\documentclass[twocolumn]{article}
\usepackage[utf8]{inputenc}

\usepackage{amsthm}
\usepackage{amssymb}
\usepackage{mathtools}
\usepackage{braket}
\usepackage{xfrac}
\usepackage{listings}

\usepackage{xcolor}
\usepackage{graphicx}

\usepackage[style=numeric-comp,sorting=none,natbib=true]{biblatex}
\renewbibmacro{in:}{}
\AtEveryBibitem{%
  \clearfield{issn} 
  \clearfield{isbn} 
  \iffieldundef{doi}{}{\clearfield{url}} 
}
\usepackage{hyperref}
\hypersetup{colorlinks=true,urlcolor=blue}

\usepackage{authblk}

\numberwithin{equation}{section}

\usepackage{abstract}

\usepackage[page,toc,titletoc]{appendix}

\title{\vspace{-3cm}Continuous Variable Quantum Algorithms: an Introduction}

\author[1]{Samantha Buck \thanks{Email: sbuck@uoguelph.ca}}
\author[1]{Robin Coleman \thanks{Email: rcolem01@uoguelph.ca}}

\author[2]{Hayk Sargsyan \thanks{Email: sargsyan.hayk@ysu.am}}

\affil[1]{University of Guelph, 50 Stone Rd E, Guelph, ON. Canada}
\affil[2]{Department of Physics, Yerevan State University, 1 Alex Manoogian, 0025 Yerevan, Armenia}
\date{}

\begin{document}

\twocolumn[
  \maketitle
  \begin{onecolabstract}
    Quantum computing is usually associated with discrete quantum states and physical quantities possessing discrete eigenvalue spectrum. However, quantum computing in general is any computation accomplished by the exploitation of quantum properties of physical quantities, discrete or otherwise. It has been shown that physical quantities with continuous eigenvalue spectrum can be used for quantum computing as well. Currently, continuous variable quantum computing is a rapidly developing field both theoretically and experimentally. In this pedagogical introduction we present the basic theoretical concepts behind it and the tools for algorithm development. The paper targets readers with discrete quantum computing background, who are new to continuous variable quantum computing.
  \end{onecolabstract}
  \vspace{0.5cm}
]
\saythanks

\tableofcontents

\section*{Introduction}
\addcontentsline{toc}{section}{Introduction}

Nowadays, one usually encounters quantum mechanics either because they are a physics (or related) student, or they got interested in quantum computing. In any case it's extremely easy to be carried away by the beauty of discreteness all-around quantum mechanics and forget about continuous quantum phenomena. There is a clear modern trend in transforming the curricula of quantum mechanics into discrete quantum computing biased ones, where discrete phenomena and calculations in finite dimensional Hilbert spaces are emphasized. The "older" way of teaching based on coordinate representation wave functions, de Broglie waves, Fourier transforms, etc. seems to be fading away, however, this kind of knowledge is crucial for continuous variable quantum computing (CVQC). In this pedagogical introduction we aim to bridge this gap. Here we present the basic mathematical concepts and tools used when dealing with continuous quantities in quantum mechanics and explain how to use them for CVQC algorithm development.

Many physical quantities, such as position and momentum or the quadratures of electromagnetic field, can accept values from a  continuous spectrum in quantum mechanics. Due to the nature of quantum mechanics and Heisenberg's uncertainty relations the precise manipulation of continuous quantum quantities is fundamentally impossible. Moreover, the existence of noise in quantum systems further worsens the situation and it seems that there is no perspective in using continuous quantum quantities for computation. However, the developments in relevant experimental realizations and quantum error correction codes have motivated the investigation of CVQC as an independent computational paradigm. The question of universality of CVQC is a subtle one, however it has been addressed in the restricted case of polynomial hamiltonians \citep{Lloyd1999}. Since then, a number of algorithms have been developed for CVQC. Besides the algorithms originally specific to continuous variables, most of the well-known quantum algorithms in discrete quantum computing (Deutsch–Jozsa, Grover, Shor, etc.) have been adapted to the continuous variable setting.

Quantum computing, in general, is any computation which is done by the exploitation of quantum properties of physical quantities, be they discrete, continuous or the combination of both\footnote{It is possible for a physical quantity to have an eigenvalue spectrum which is discrete in some region(s) and continuous in other(s).}. Moreover, it is also possible to combine systems with continuous variables with traditional discrete quantum computing and implement the so called hybrid computing \cite{Lloyd2000, Wang2001}. Our focus in this paper is the CVQC as an independent paradigm.

Section \ref{sec:quantum_continuous_variables} will serve as a crash course on relevant concepts and definitions known from quantum mechanics. Explanations go as deep as necessary for the overall integrity and scope of the text. For more detailed explanations from ground up the reader is invited to read quantum mechanics textbooks (e.g. \cite{Griffiths, CohenTannoudji, Landau1977}), however it is not necessary for complete understanding of this paper.

Sections \ref{sec:hilbert_space} and \ref{sec:quantum_harmonic_oscillator} are incremental development on top of the previous section to cover the notion of Hilbert spaces in relation to continuous variables, and quantum harmonic oscillator. These are crucial topics for CVQC and algorithm development. 

In section \ref{sec:cv_algorithms} we discuss a few famous algorithms to demonstrate how all the mathematical tools introduced in the previous section work out in real life. We also briefly introduce \textit{strawberryfields} \cite{Killoran2019}, a Python library which can be used for CVQC simulation. 

Section \ref{sec:summary} summarizes the paper and provides some future directions.

\section{Quantum Continuous Variables} \label{sec:quantum_continuous_variables}

In the quantum computing literature the term \textit{continuous variable} is loosely used to refer to a physical quantity that can accept any value in an interval - in contrast to discrete variables which can accept only distinct values. An example of a continuous variable is the \(x\) coordinate of a particle. In quantum mechanics physical quantities are represented in terms of operators\footnote{In this paper we have adopted the standard convention of denoting an operator with a hat on top, e.g. \(\hat{x}\)} and their continuity properties are encoded in the eigenvalue spectrum of these operators. If the eigenvalue spectrum of some operator is continuous then the corresponding physical quantity can serve as a continuous variable for computation. In this paper we use the term \textit{continuous variable} as a collective reference to the physical quantity itself, the corresponding operator and the eigenvalue.

\subsection{General Concepts}

Any continuous variable has a conjugate which is yet another continuous variable. In the case of the coordinate \(\hat{x}\) the conjugate quantity is the \(x\) component of momentum \(\hat{p}_x\). For simplicity we will drop the subscript of the momentum operator. In this paper we will almost always denote a continuous variable and its conjugate by \(\hat{x}\) and \(\hat{p}\), however this does not necessarily mean they are coordinate and momentum. They can be any pair of conjugate continuous variables. We just use this notation to continuously hint that the mathematical properties of these variables are exactly the same as for coordinate and momentum. The physical realizations can be non-related to coordinate and momentum. 

\(\hat{x}\) and \(\hat{p}\) satisfy the following commutation relation
\begin{equation} \label{eq:commutation}
[\hat{x}, \hat{p}] = i \hbar ,
\end{equation}
where \([\hat{x}, \hat{p}] = \hat{x}\hat{p} - \hat{p}\hat{x}\) is the commutator. As you can see the two conjugate continuous variables do not commute. The result of this non-commutativity is another important property, which is the Heisenberg's uncertainty principle
\begin{equation}
\label{eq:uncertainty_relation}
\Delta x\Delta p \ge \frac{\hbar}{2},
\end{equation} 
where \(\Delta x = \sqrt{\overline{(\hat{x}-\bar{x})^2}}\) is the standard deviation, and the overbar notation means ensemble average. If this is completely new to you, appendix \ref{ap:A} provides more details and explanation about \eqref{eq:uncertainty_relation}. 

As we know the state of a system in quantum mechanics is described by a wave function\footnote{We consider only pure states, so we don't talk about density matrices, only wave functions.}. A wave function is the amplitude of a probability distribution, i.e. the absolute squared of a wave function gives the probability distribution over the values of its argument. Consider a quantum system described by a wave function \(\psi (q)\) where \(q\) is a continuous variable. In this case it is said that the wave function is in the \(q\) representation. In particular the quantity \(\vert \psi(q) \vert^2 dq\) is the probability for the system to have value of \(q\) in the interval \(q, q+dq\). $q$ is oftentimes referred to as the "space" one is working in. In practise, \(q\) can be any physical quantity, for example coordinate $x$, momentum $p$, energy $E$, etc. The same state of a system can be described in different representations and there should be some relation between the wave functions in different representations. These relations give transformation laws, so that having the wavefunction in one representation one can transform it into the wavefunction in another representation. It is known from quantum mechanics that the transformation law is particularly straightforward if one transforms from a representation to its conjugate representation: it is a Fourier Transform. Given a state \(\phi(p)\) in the \(p\) representation the corresponding state in \(x\) representation can be obtained via the Fourier transform
\begin{equation} \label{eq:p_to_x_transform}
\psi(x)=\frac{1}{\sqrt{2 \pi \hbar}} \int_{-\infty}^{\infty} \phi(p) e^{i p x / \hbar} d p 
\end{equation}\\
and vice versa
\begin{equation} \label{eq:x_to_p_transform}
\phi(p)=\frac{1}{\sqrt{2 \pi \hbar}} \int_{-\infty}^{\infty} \psi(x) e^{-i p x / \hbar} d x
\end{equation}

When we choose a particular representation, we can also talk about the explicit forms of the operators \(\hat{x}\) and \(\hat{p}\). In the \(x\)-representation the operator \(\hat{x}\) just coincides with its numerical value \(x\) and the operator \(\hat{p}\) has the following explicit form
\begin{equation} \label{eq:p_operator}
\hat{p} = -i\hbar\frac{\partial}{\partial x}. 
\end{equation}
Conversely, in the \(p\)-representation the operator \(\hat{p}\) just coincides with its numerical value \(p\) and the operator \(\hat{x}\) has the following explicit form
\begin{equation} \label{eq:x_operator}
\hat{x} = i\hbar\frac{\partial}{\partial p}. 
\end{equation}
The explicit forms of the operators \(\hat{x}\) and \(\hat{p}\) in any representation \(q\) can be derived based on the above formulas, but we will not cover it here since it's outside of the scope of the current paper. The commutation relation \eqref{eq:commutation} and the uncertainty principle \eqref{eq:uncertainty_relation} are always the same, regardless of the representation.

\subsection{Eigenvalues and eigenfunctions}

Given a continuous variable \(\hat{f}\), the eigenvalue equation in some \(q\)-representation is
\begin{equation} \label{eq:eigenvalue_equation}
\hat{f}\psi_f (q) = f\psi_f (q).
\end{equation}
In the discrete case we usually enumerate the eigenvalues and eigenvectors with integer indices. In the continuous case we just use the letter without a hat to denote the eigenvalues and the same letter as a subscript to denote the corresponding eigenstate. The eigenstates \(\psi_f (q)\) describe a state where the physical quantity underlying the operator \(\hat{f}\) has exactly determined value equal to \(f\).

It is known from quantum mechanics that an arbitrary wave function \(\psi(q)\) can be decomposed (i.e. written as a superposition) in terms of eigenstates of any continuous variable \(\hat{f}\) as follows

\begin{equation} \label{eq:wave_function_expansion}
\psi (q) = \int_{-\infty}^{+\infty}c(f)\psi_f(q)df,
\end{equation}
where the function \(c(f)\) represents the decomposition coefficients. Recall, that in the case of discrete variables we have a summation over discrete eigenstates, and a set of coefficients \(c_{n}\). In the continuous case the summation transforms into integration, and the set of coefficients becomes a continuous function \(c(f)\).  The function \(c(f)\) is the wavefunction of the state under consideration in the \(f\)-representation and the equation \eqref{eq:wave_function_expansion} gives transformation law between any two representations \(f\) and \(q\). This is easily seen in the particular case of \(\hat{q}=\hat{x}\) and \(\hat{f}=\hat{p}\). Given the formulas \eqref{eq:eigenvalue_equation} and \eqref{eq:p_operator} the eigenfunctions of \(\hat{p}\) in the \(x\)-representation are determined from the following equation
\begin{equation} \label{eq:eig_momentum}
-i\hbar\frac{\partial \psi_{p}(x)}{\partial x} = p\psi_{p}(x)
\end{equation}
and they are equal to\footnote{The solution for the equation \eqref{eq:eig_momentum} alone is \(\psi_{p}(x) = \text{const}\times e^{\frac{i}{\hbar}p x}\). The value of the constant is determined from normalization condition which will be covered in the next subsection.} 
$$
\psi_{p}(x) = \frac{1}{\sqrt{2\pi\hbar}} e^{\frac{i}{\hbar}p x}.
$$
Substituting this into the equation \eqref{eq:wave_function_expansion} we instantly recover the Fourier transformation law \eqref{eq:p_to_x_transform}. Conversely, considering \(\hat{q}=\hat{p}\) and \(\hat{f}=\hat{x}\) the inverse Fourier transform \eqref{eq:x_to_p_transform} is recovered.

In the theory of CVQC we are quite often interested in states with exactly determined value of \(\hat{x}\) or \(\hat{p}\) (i.e. eigenstates of \(\hat{x}\) or \(\hat{p}\)). If in \eqref{eq:wave_function_expansion} we take \(\hat{f}=\hat{p}\), we obtain the expansion of any \(q\)-representation wave function in terms of eigenstates of \(\hat{p}\) (i.e. in terms of states where \(\hat{p}\) has exactly determined value)

\begin{equation}
\psi (q) = \int_{-\infty}^{+\infty}c(p)\psi_{p}(q)dp
\end{equation}
Let's consider the expansion of the wave function for a particular state - the state \(\psi_x (q)\) which is an eigenstates of \(\hat{x}\) in the \(q\)-representation (i.e. has exactly determined coordinate \(x\)). We get

\begin{equation}
\psi_x (q) = \int_{-\infty}^{+\infty}c_x(p)\psi_{p}(q)dp,
\end{equation}
where \(c_x(p)\) is now the wave function of the state with exact coordinate \(x\) in the \(p\)-representation. Given the definition \eqref{eq:x_operator}, we know that the eigenstates \(c_x(p)\) are determined from the eigenvalue equation \(i\hbar\frac{\partial c_x(p)}{\partial p} = xc_x(p)\) and are equal to \(c_x(p)=\frac{1}{\sqrt{2\pi\hbar}}e^{-\frac{i}{\hbar}px}\). So we have the following relation

\begin{equation} \label{eq:psi_x_q}
\psi_x (q) = \frac{1}{\sqrt{2\pi\hbar}}\int_{-\infty}^{+\infty}e^{-\frac{i}{\hbar}px}\psi_p(q)dp
\end{equation}

With analogous considerations the reverse formula can be obtained as well

\begin{equation} \label{eq:psi_p_q}
\psi_p (q) = \frac{1}{\sqrt{2\pi\hbar}}\int_{-\infty}^{+\infty}e^{\frac{i}{\hbar}px}\psi_x(q)dx
\end{equation}

\subsection{Bra-ket notation}
In discrete quantum computing we are used to Dirac's bra-ket notation. In CVQC, the bra-ket notation can be used as well, however there are a few caveats and the bra-ket notation should be used with caution.

In the previous subsections we have mostly used \(q\) as placeholder representation whenever we didn't want to concretize the representation and derive general formulas regardless of the representation. In the bra-ket notation a state with exact coordinate \(x=x_0\), and not concretized representation is denoted as \(\vert x_0 \rangle\).
A state with exact momentum \(p=p_0\), and not concretized representation is denoted as \(\vert p_0 \rangle\).
Whenever we want to concretize the representation, it is specified as a subscript to the bra-kets, (e.g. \(\vert x_0 \rangle_p\)). With this notation, the formulas \eqref{eq:psi_x_q} and \eqref{eq:psi_p_q} have the following form

\begin{align} \label{xket}
\vert x \rangle &= \frac{1}{\sqrt{2\pi\hbar}}\int_{-\infty}^{+\infty}e^{-\frac{i}{\hbar}px} \vert p \rangle dp \\
\vert p \rangle &= \frac{1}{\sqrt{2\pi\hbar}}\int_{-\infty}^{+\infty}e^{\frac{i}{\hbar}px} \vert x \rangle dx 
\end{align}

The bra vectors are defined as complex conjugate of the ket vectors. The inner product of a bra and a ket is defined as follows
\begin{equation}
    \langle x_1 \vert x_2 \rangle = \int_{-\infty}^{+\infty}\psi_{x_1}^{\ast}(q)\psi_{x_2}(q)dq .
\end{equation}
An important caveat compared to the case of discrete variables is the normalization of bra-kets. From discrete variable quantum computing we know that two eigenkets corresponding to different eigenvalues of some operator are orthogonal (i.e. their inner product is equal to zero), and each eigenket is normalized to \(1\) (i.e. the inner product of a ket with itself is equal to \(1\)).  If there were a discrete set of possible values for \(x\), denoted as \(x_{i}\), we could simply adopt the Kronecker delta notation to summarize this orthonormalization conditions in one equation \(\left\langle x_{i} \mid x_{j}\right\rangle=\delta_{x_{i} x_{j}}\). This is not true in the continuous case, and it can be easily seen for example by calculating the quantity \(\langle p\vert p \rangle\) in the \(x\)-representation

\begin{align}
    \langle p\vert p \rangle &= \int_{-\infty}^{+\infty} \psi_{p}^{\ast}(x)\psi_{p}(x)dx \notag \\
    &= \frac{1}{2\pi\hbar} \int_{-\infty}^{+\infty}dx. \notag
\end{align}
It is readily seen that this expression is nowhere near being equal to \(1\), it is in fact infinite. In quantum mechanics this problem is solved by introducing Dirac's delta function instead of the Kronecker's notation 
\begin{equation} \label{eq:orthonormalization}
\left\langle x^{\prime} \mid x\right\rangle=\delta\left(x^{\prime}-x\right)
\end{equation}
where \(\delta\left(x^{\prime}-x\right)\) is the Dirac delta function. If you are not familiar with Dirac's delta function, appendix \ref{ap:B} contains additional explanations and useful formulas for you. The fact that a ket \(\vert x \rangle\) is not normalizable to \(1\) is sometimes referred to as "non square integrability", meaning that the integral of wave function absolute squared is not a finite number. There are not physical states corresponding to these non-normalizable kets - the kets of real states are always normalizeable to \(1\). From quantum mechanics we know that the normalization integral is an integral over a probability distribution, and the integration over the entire range should always be \(1\). Nevertheless, in quantum mechanics it is allowed to use these states as mathematical tools and they make a lot of calculations easier. One just needs to keep in mind and always use the correct orthonormalization relation \eqref{eq:orthonormalization}.

\section{Hilbert Space}
\label{sec:hilbert_space}

In discrete quantum computing one uses the eigenstates of a discrete variable (e.g. the hamiltonian, in which case the eigenstates are the discrete energy levels) as computational basis. In the case of CVQC we would like to be able to act analogously, (i.e. use the eigenstates of a continuous variable as a computational basis). It is well known that for discrete quantum computing it is convenient to describe computations in an abstract Hilbert space \cite{NielsenChuang}. CVQC is not any different, however the topic needs cautious handling. The concept of a Hilbert space in the discrete case is pretty straightforward (see e.g. \cite{VlatkoVedral} for a simple introduction). In layman's terms, it can be thought of as a space with a lot of perpendicular axes and each discrete eigenstate of a quantum system is considered as a unit vector along one of the axes. Any state of the system is represented as a vector and can be understood as a linear combination (superposition) of such unit vectors. If one tries to do the same for eigenstates corresponding to a continuous variable they will immediately notice that the Hilbert space is infinite dimensional\footnote{This might not seem an indicative of a problem at first glance, since infinite dimensional Hilbert spaces are abundant in quantum mechanics even for discrete variables. However, for discrete quantum computational purposes the Hilbert space is always finite dimensional.} and the labeling of axes is continuous. This means the axes are not countable, i.e. one cannot actually index them.

To demonstrate the above on an example, let's consider a spin-\sfrac{1}{2} particle which is a typical representative of a qubit. One could use a 2-dimensional Hilbert space for this system, with two basis kets, \(|up\rangle\) and \(|down\rangle\). The wavefunction of an arbitrary state would be a superposition of these two basis kets and we would talk about probabilities \(P(+\sfrac{1}{2})\) and \(P(-\sfrac{1}{2})\) for the system to be \(|up\rangle\) and \(|down\rangle\) respectively. We don't talk about probabilities such as \(P(0.3)\) or \(P(0.301)\), since there are no corresponding states in our basis. However, an observable like position is continuous, where we \textit{can} ask the probability that the position is 0.3, or 0.301, etc. Thus the kets corresponding to each possible value of the coordinate should be present in the basis of the Hilbert space. Since there are infinitely many  possible values of position, we need infinitely many axes in the Hilbert space. Moreover, since the possible values of coordinate are not countable (i.e. for every two distinct values one can always find another one which is in between), the axes in the Hilbert space are not countable either.

Yet another problem with this approach of constructing a Hilbert space is the fact that the basis kets are not normalizable (see section \ref{sec:quantum_continuous_variables} for more details). However, the mathematical apparatus of quantum mechanics allows having the construction of Hilbert spaces with non-physical basis kets \cite{CohenTannoudji}, so we use the Hilbert space construction described above for CVQC.

A continuous variable in quantum mechanics can accept values in the infinite range $(-\infty, +\infty)$ or some finite range $[a, b]$. The formalism is the same for both cases, so we will just consider the case of infinite range. Similar to the case of discrete quantum computing we can talk about superpositions of basis ket vectors and entanglement between variables. A superposition state is characterized by a complex function \(f(x)\) representing the amplitude of each ket

\begin{equation} \label{eq:superposition}
\vert \psi \rangle=\int_{-\infty}^{+\infty}f(x)\vert x \rangle dx.
\end{equation}
Recall that in the case of discrete states one has summation over the basis and here we have integration. A simple example of an entanglement can be represented by a function \(g(x)\) and a real constant \(c\)
\begin{equation}
\vert \psi \rangle = \int_{-\infty}^{+\infty}g(x)\vert x, x-c \rangle.
\end{equation}
In this example the first and the second registers are perfectly correlated, so that if, for example, the first one is measured to be some value \(x_0\), the second one is immediately known to be \(x_0 - c\).

We can also talk about various operations (or gates) acting on states in this Hilbert space. By far the most common one is the Fourier gate \(\mathcal{F}\) which is the analog of the Hadamard gate.  The Fourier transform takes a ket \(\vert x \rangle\) to a ket \(\vert p\rangle\) of the conjugate variable: \(\mathcal{F}\vert x\rangle = \vert p\rangle\). Thus, its action can be explicitly written based on eq. \eqref{xket}
$$
\mathcal{F}\vert x\rangle=\frac{1}{\sqrt{2\pi \hbar}} \int e^{\frac{i}{\hbar} x y}\vert y\rangle \mathrm{d} y 
$$
where both \(x\) and \(y\) are position kets.

Another common and simple operator in CVQC is the displacement operator. In its most general form it can be represented as an operator depending on two parameters \(\Delta x\) and \(\Delta p\)

\begin{equation}
\hat{D}(\Delta x, \Delta p) = e^{i\Delta x \hat{p}-i\Delta p \hat{x}}.
\end{equation}
This operator in essence displaces the state in the position space by \(\Delta x\) and in momentum space by \(\Delta p\). When multiple displacement operators are applied it can be shown that it can be equivalently described by a single displacement operator with a euclidean sum of the displacements and the momenta of the constituent operators \cite{PotocekDisplacement}. The displacement operator and other operators will be discussed in greater detail within the context of the quantum harmonic oscillator in section \ref{sec:quantum_harmonic_oscillator} .

Let's also briefly discuss a projective measurement on value $x_0$. One might naively think, that in analogy with the discrete case, this projection operation can be represented as $\vert x_0 \rangle \langle x_0 \vert$, however this does not work for continuous variables. There are two main reasons for this:
\begin{enumerate}
    \item projection operation defined this way is not idempotent\footnote{Idempotent means that if you apply the operator multiple times one immediately after another, you get the same effect as if you applied the operator only once.}.
    \item no measuring apparatus can have infinite precision to exactly distinguish the value $x_0$ from its arbitrarily close neighbors.
\end{enumerate}
Both of these problems are resolved if we consider a finite precision projection operation
\begin{equation} \label{eq:measurement_def}
    \hat{P}_{x_0} = \int_{x_0-\Delta x/2}^{x_0+\Delta x/2} \vert x \rangle\langle x \vert dx,
\end{equation}
where $\Delta x$ is the precision. With this operator, the second concern resolved by definition. Let us convince ourselves that this operator is indeed idempotent and the first concern is resolved as well. For this purpose it is enough to show that the action of this operator on an arbitrary superposition state \(\vert \psi\rangle\) is the same as its double action on the same state, i.e. \(\hat{P}_{x_0}\hat{P}_{x_0}\vert \psi\rangle=\hat{P}_{x_0}\vert\psi\rangle\). Indeed 
\begin{align}
    \hat{P}_{x_0}\vert\psi\rangle &= \int_{x_0-\Delta x/2}^{x_0+\Delta x/2} \vert x \rangle\langle x \vert dx \int_{-\infty}^{+\infty}f(x^{\prime})\vert x^{\prime} \rangle dx^{\prime} \notag \\
    &= \int_{x_0-\Delta x/2}^{x_0+\Delta x/2}\vert x \rangle dx\int_{-\infty}^{+\infty}f(x^{\prime}) \langle x \vert x^{\prime} \rangle dx^{\prime} \notag \\ 
    &= \int_{x_0-\Delta x/2}^{x_0+\Delta x/2}\vert x \rangle dx\int_{-\infty}^{+\infty}f(x^{\prime})   \delta(x-x^{\prime}) dx^{\prime} \notag \\
    &=  \int_{x_0-\Delta x/2}^{x_0+\Delta x/2}f(x) \vert x \rangle dx
\end{align}
and
\begin{align}
    \hat{P}_{x_0}\hat{P}_{x_0}\vert \psi\rangle &= \int_{x_0-\Delta x/2}^{x_0+\Delta x/2} \vert x \rangle\langle x \vert dx \int_{x_0-\Delta x/2}^{x_0+\Delta x/2}f(x^{\prime})\vert x^{\prime} \rangle dx^{\prime} \notag \\
    &= \int_{x_0-\Delta x/2}^{x_0+\Delta x/2}f(x) \vert x \rangle dx
\end{align}

\section{Quantum Harmonic Oscillator}
\label{sec:quantum_harmonic_oscillator}

So far we have considered the properties of continuous variables in quantum mechanics in general, without assuming a particular physical system. There are several physical systems which can be used to accomplish CVQC, however the most famous is the implementation with quantum optical systems. The continuous variables in these systems are the quadratures of the electromagnetic field, however one does not need knowledge in quantum electrodynamics in order to study the continuous variable algorithms. The mathematical description of the quadratures in a quantum optical mode is exactly the same as the mathematical description of a quantum harmonic oscillator.

A one dimensional quantum harmonic oscillator is described by the following Hamiltonian
\begin{equation}
\hat{H} = \frac{\hat{p}^2}{2} + \frac{\hat{x}^2}{2} = -\hbar^2\frac{\mathrm{d}^2}{\mathrm{d}x^2} + \frac{x^2}{2}.
\end{equation}
We have taken the mass and the frequency of the oscillator to be \(1\). As a result the reduced Planck constant \(\hbar\) is dimensionless, however we do not assign any particular numerical value to it in this paper and we keep it as a letter throughout\footnote{In the literature there are several conventions about assigning a value to \(\hbar\), e.g. some people prefer to set \(\hbar=1\), others set it as \(\hbar=2\) or even \(\hbar=1/2\). We keep \(\hbar\) as a symbol in this paper to be consistent with everyone.}.
We don't need to care about the frequency either, since in usual realizations of CVQC frequencies of all oscillators are the same.

Next, let's define the following complex operator
\begin{equation}
\hat{a} = \frac{1}{\sqrt{2\hbar}}(\hat{x}+i\hat{p}).
\end{equation}
The coordinate and the momentum operators are given in terms of \(\hat{a}\) and \(\hat{a}^{\dagger}\) as follows
\begin{gather}
\hat{x}=\sqrt{\frac{\hbar}{2}}(\hat{a}+\hat{a}^{\dagger}) \\
\hat{p}=-i\sqrt{\frac{\hbar}{2}}(\hat{a}-\hat{a}^{\dagger}).
\end{gather}
If we then have the operator $\hat{a}^{\dagger}$ or $\hat{a}$ act on the energy eigenstate of the system, $\ket{n}$, we see that it produces a higher or lower eigenstate respectively
\begin{gather}
    \hat{a}^{\dagger} \ket{n} = \sqrt{n+1}\ket{n+1} \label{creation}\\
    \hat{a} \ket{n} = \sqrt{n}\ket{n-1} \label{annihilation}\quad . 
\end{gather}
This means that  $\hat{a}^{\dagger}$ essentially adds a quantum of energy to the system, hence this operator is generally referred to as the creation operator. The opposite is true for  $\hat{a}$, leading to the name, annihilation operator. We can also combine these to get the quanta of energy out of an eigenstate, this is called the number operator $\hat{N}$ given as
\begin{gather}
    \hat{N} = \hat{a}^\dagger \hat{a} \\
    \hat{N} \ket{n} = n \ket{n}
\end{gather}
We can also use these operators to define some useful operations, namely the phase shift and displacement operators. These are defined as $\hat{U}(\theta)$ and $\hat{D}(\alpha)$ respectively and represent the following:
\begin{gather}
    \hat{U}(\theta) = e^{-i\theta \hat{N}} \\
    \hat{D}(\alpha) = e^{\alpha \hat{a}^\dagger - \alpha^* \hat{a}}.
\end{gather}
Before we start to construct states with these operators we need to consider how to visualize these states. One useful measure that we use is the Wigner quasi-probability distribution (or Wigner function) over the phase space (i.e. the joint space of coordinate and momentum). 

\begin{figure}
    \centering
    \includegraphics[width=8cm]{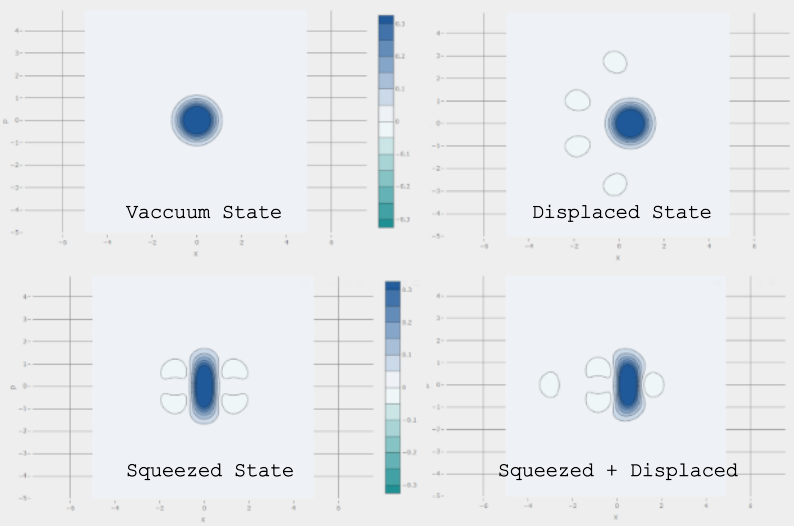}
    \caption{These contour plots show the Wigner function values in phase space for a vaccuum state (top left), a state displaced with the parameter of $0.5$ (top right), a vacuum state squeezed with the parameter $0.5$ (bottom left), and a vacuum state with both the aforementioned displacement and squeezing applied (bottom right).}
    \label{fig:WignerStates}
\end{figure}

With the help of the aforementioned operators we can define coherent states, which mathematically are eigenstates of the annihilation operator, herefore
\begin{equation}
    \hat{a}\ket{\alpha} = \alpha \ket{\alpha}
\end{equation}
These coherent states inherently have circular uncertainly in phase space due to the uncertainty principle. These states can then be squeezed, in that a squeezing operator can act on the coherent state to squeeze the circular uncertainty into an ellipse in phase space. Mathematically this squeezing operator is written as
\begin{equation}
    \hat{S}(\zeta) = e^{1/2(\zeta^* \hat{a}^2 - \zeta \hat{a}^{\dagger2})}
\end{equation}
We can now create coherent squeezed states starting from vacuum states by simply applying the displacement operator and the squeezing operator. Xanadu's strawberryfields offers a comprehensive quantum computing toolkit that can be used in python or on their website in an interactive applet. This program comes with all of the aforementioned operations to prepare these states to be used in algorithms that will be discussed in detail in section \ref{sec:cv_algorithms}. We can observe how these states change by plotting the Wigner function. As an example, a vacuum state can be created and then displacement and squeezing operators act on it to obtain a displaced squeezed state. With Dgate applying the displacement operator and Sgate applying the squeezing operator, they can be done together in one line with the DisplacedSqueezed function.
\begin{lstlisting}[language=python]
import strawberryfields as sf
from strawberryfields.ops import *

prog = sf.Program(1)
with prog.context as q:
    # The initial states are 
    # prepared with q[0] 
    Vacuum() | q[0]
    Dgate(1) | q[0]
    Sgate(1) | q[0]
    # Alternatively this can be
    # Done with the following:
    # DisplacedSqueezed(1,0,1,0) | q[0]
\end{lstlisting}
and we can see how the Wigner function changes with the three steps in Figure \ref{fig:WignerStates}.

All of the operators introduced above act on a single harmonic oscillator, but systems can have multiple coupled harmonic oscillators. These coupled oscillators can interact through various operators. One such operator is the mixing\footnote{In optics this is the beamsplitter, however since in this paper we are avoiding optics as much as we can, we use the term "mixing" instead} which is defined as 
\begin{equation}
    \hat{B}(\theta, \phi) = \exp{\left(\theta\left(e^{i \phi}\hat{a}_1\hat{a}_2^\dagger -e^{i \phi}\hat{a}_1^\dagger\hat{a}_2 \right)\right)}
\end{equation}
where $\theta$ represents the transmittivity angle between the and $\phi$ is the phase angle. We can create a $50-50$ mixer by setting $\theta = \pi / 4$, and the mixer will be symmetric by setting $\theta = \pi / 2$. 

At this point we can now create harmonic oscillators in coherent states representing laser light starting from the vacuum state and we can begin to couple these systems. We have thus far, however, introduced continuous variables in the context of a real optical system without discussing how these continuous variables are handled in a computational sense. So let's take a moment to discuss these continuous variables used in computation.

\section{Continuous Variable Quantum Algorithms}
\label{sec:cv_algorithms}

In this section we take a look at several quantum algorithms to give the reader an idea of how all the introduced mathematical concepts work. There are many algorithms for continuous variable quantum computing, and they can be grouped into two categories: 1. algorithms originally specific to continuous variables, 2. algorithms originally invented for discrete variable quantum computing and then adapted for continuous variables. We will restrict ourselves in looking at a few examples from the second category only.

As we have already mentioned in the sections \ref{sec:quantum_continuous_variables} and \ref{sec:hilbert_space} the exact eigenstates \(\vert x \rangle\) of a continuous variable \(\hat{x}\) are not physical states, hence they are not realizable in any real physical device. In reality you can create highly localized states around some value \(x\) but never an exactly localized one. Nevertheless, on the theoretical front, the mathematical apparatus of quantum mechanics provides tools to work with these states without creating any theoretical artifacts. Being able to work in a Hilbert space defined by these non-physical states makes things a lot easier. When developing CVQC algorithms one usually goes through the following procedure:
\begin{enumerate}
    \item Develop the algorithm assuming that exact eigenstates are physically possible. 
    \item If applicable, investigate the features and properties of the algorithm and the speedup over the classical counterpart.
    \item Drop the assumption about exact eigenstates and investigate the algorithm by replacing them with realistic states. A realistic state corresponding to an exact eigenstate \(\vert x_0 \rangle\) is a state which is localized around \(x_0\), so can be described as a continuous superposition of exact eigenstates as defined in \eqref{eq:superposition}. A common choice for realistic states (which is also easy to achieve experimentally) are states with Gaussian distribution \ref{sec:quantum_harmonic_oscillator}. 
    \begin{equation}
        \vert \psi \rangle = \frac{\sqrt{R}}{\pi^{1/4}} \int_{-\infty}^{+\infty} e^{-\frac{R^2}{2}x^2} \vert x \rangle dx. \notag
    \end{equation}
    For the conjugate variable \(p\) this state is Gaussian as well
    \begin{equation}
        \vert \psi \rangle = \frac{1}{\sqrt{R}\pi^{1/4}} \int_{-\infty}^{+\infty} e^{-\frac{1}{2R^2}p^2} \vert p. \rangle dp \notag
    \end{equation}
    These are exactly the squeezed states introduced in the section \ref{sec:quantum_harmonic_oscillator}, and \(R>1\) is the squeezing factor. Note that if \(R>1\), the state is squeezed in the \(x\) direction and stretched in the \(p\) direction, and vice versa if \(R<1\). The fact that the states cannot be squeezed in both directions is the direct consequence of Heisenberg's uncertainty relations. A more complete introduction to squeezed states from the perspective of continuous variables can be found in \cite{Lvovsky2016}.
\end{enumerate}

\subsection{Grover's algorithm.} 

In Grover's quantum search algorithm, an unsorted list of \(N\) entries can be searched to find an unmarked item with \(O(\sqrt{N})\) steps, instead of the best classical case which requires \(O(N)\) steps. This quantum search algorithm, originally proposed for discrete-variable quantum systems, can be generalised to the continuous variable representation \cite{Pati2000Grover}.

From the theory of discrete quantum computing we know that the basic elements required for Grover's search algorithm with discrete-variables are an oracle which marks the solution to the searching problem, Hadamard gates to generate state superposition, and finally the Grover diffusion operator, which is also sometimes referred to as an amplitude amplification operation. The oracle is given as a black box operation and we are not aware of its internal workings. What we know is the way the oracle marks the special element: given a superposition of all possible states the oracle inverts the sign of the special state. This is usually done in two equivalent ways: 1. using an ancillary qubit to handle the sign flip based on the oracle's output, and 2. in-place sign inversion without using a special ancillary qubit. When considering the continuous variable version of Grover's algorithm below, we will concentrate on the in-place variant.

The statement of Grover's search problem in the continuous variable setting can be understood as follows. There is a continuous variable \(x\) and an oracle which highlights a special value \(x_f\) from the interval \((-L/2, L/2)\) with some precision \(\Delta x\). Essentially, we have a collection of \(N=\frac{L}{\Delta x}\) intervals and our task is to find the one which contains the value \(x_f\). The continuous variable Grover algorithms achieves this in \(O(\sqrt{N})\) steps.

Consider a continuous variable in the initial state \(\vert x_{0}\rangle\) and apply the Fourier transform on it\footnote{The original paper on continuous variable Grover search uses the notation \(\hbar = \frac{1}{2}\) and considers \(n\) continuous variables. For the sake of simplicity we consider searching along one continuous variable only. The Fourier transform in the case of \(n\) variables is 
$$
\mathcal{F}\vert x\rangle=\frac{1}{\sqrt{(2\pi \hbar)^n}} \int  e^{\frac{i}{\hbar} x y}\vert y\rangle \mathrm{d} y
$$}
$$
\mathcal{F}\vert x_{0}\rangle=\frac{1}{\sqrt{2\pi \hbar}} \int e^{\frac{i}{\hbar} x_{0} x}\vert x\rangle \mathrm{d} x
$$
Then apply the oracle \(\mathcal{O}\). Next, apply an inverse Fourier transform followed by a selective inversion operator about the state \(\vert x_0\rangle\) 
$$
I_{x_0} \mathcal{F}^{\dagger} \mathcal{O}
$$
where \(I_{x_0} = 2 \hat{P}_{x_0} - \mathbb{I}\) is the inversion operation, with \(\mathbb{I}\) being the identity operator and \(\hat{P}_{x_0}\) the projection operator defined in \eqref{eq:measurement_def}. 

Grover's algorithm then proceeds by repeating the above steps \(\sqrt{N}\) times 
$$
\dots \underbrace{I_{x_0} \mathcal{F}^{\dagger} \mathcal{O}\mathcal{F}}_\text{2}\underbrace{I_{x_0} \mathcal{F}^{\dagger} \mathcal{O} \mathcal{F}}_{1} \vert x_0 \rangle
$$
to gradually transform the initial state \(\vert x_0\rangle\) into a state where \(\vert x_f\rangle\) and its immediate surroundings have high probability. 

The operation \(\mathcal{O}\) is given as a black box oracle, so from the perspective of Grover's algorithm the implementation details and the inner workings of the oracle are not important to us. Nevertheless, if you are interested in implementing oracles, then it's worth to mention that the oracle itself can be implemented simply as an inversion operation about the state \(\vert x_f\rangle\). The original paper on continuous variable Grover search actually assumes that the oracle is given as the inversion operator \(I_{x_f}\).

The construction above assumes the existence the exact eigenstate \(\vert x_0\rangle\). To get rid of this assumption, we can substitute \(\vert x_0\rangle\) with a Gaussian distribution peaked at \(x_0\). It has been shown that with this substitution the algorithm described above still achieves success in the same complexity \cite{Pati2000Grover}.

\subsection{The Deutsch–Jozsa algorithm} 

Assuming the existence of exact eigenstates, the continuous variable Deutsch-Jozsa algorithm has been introduced in \cite{Pati2003DeutschJozsa} and the version with realistic Gaussian states has been developed in \cite{Wagner2012}. Here we give a brief overview of the results. 

Let's consider a binary function $f(x)$ where $x$ is a real number\footnote{A function mapping real numbers to binary: \(R \rightarrow\{0,1\}\)}. For simplicity let's also consider that \(x \in (-\infty, +\infty)\). The function $f(x)$ is guaranteed to be either constant or balanced. The case of a constant function is straightforward to imagine. The challenging part is to come up with a good definition of what it means for a continuous function to be balanced. The strict mathematical definition can be found in \cite{Pati2003DeutschJozsa}, but the basic idea is that if we combine all intervals where the function is equal to \(1\) the result should be of the same order as the combined interval where the function is equal to \(0\).

The function \(f(x)\) is given as an oracle and the Deutsch–Jozsa problem asks to find out whether a given oracle implements a constant function or a balanced one. This problem which would classically require infinite evaluations of the function can be solved with one evaluation using the Deutsch-Jozsa algorithm in a perfect continuous variable quantum computer (i.e. a thought computer where the exact eigenstates are physical) \cite{Pati2003DeutschJozsa}. It has been also shown that considering realistic Gaussian states instead of the idealized eigenstates, the continuous variable Deutsch-Jozsa algorithm can still be accomplished in a finite number of steps, thus still providing infinite speedup over the classical counterparts \cite{Wagner2012}.

\begin{figure}[ht]
    \centering
    \includegraphics[width=8cm]{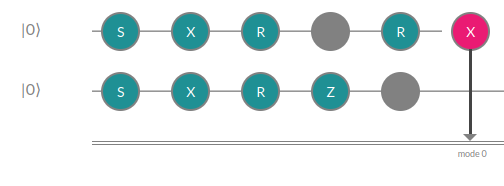}
    \caption{Continuous Variable Deutsch-Jozsa algorithm circuit visualized in Strawberry Fields. Where an X gate is used as the example oracle.}
    \label{fig:DJCircuit}
\end{figure}

\begin{figure}[ht]
    \centering
    \includegraphics[width=8cm]{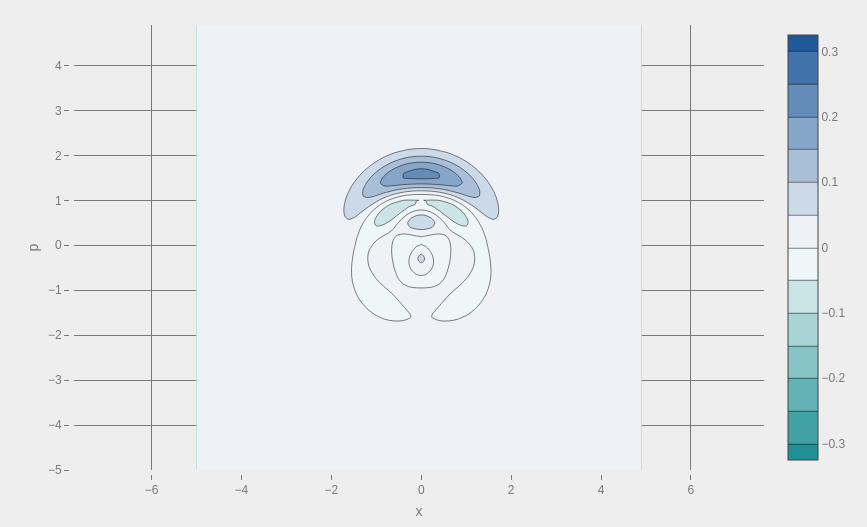}
    \caption{The second mode of the Wigner function represented as a contour plot with an initial state of $r=1.0$ and $\phi=0.5$}
    \label{fig:DJWigner}
\end{figure}

In order to accomplish this two continuous variables are prepared in the initial states and then a Hadamard gate is effectively applied through the use of the Xgate and Rgate functions available in Strawberry Fields. This is then passed to the oracle. Following the oracle another rotatition is applied to the first qunat. The first variable can then be measured and if the output is zero then the oracle function is constant and if it is non-zero then the function is balanced. This can be done in python as follows using the Xgate as a sample oracle for this situation.
\begin{lstlisting}[language=python]
import strawberryfields as sf
from strawberryfields.ops import *

prog = sf.Program(2)
with program.context as q:
    # Prepare initial states
    Squeezed(2) | q[0]
    Squeezed(2) | q[1]
    
    Xgate(3) | q[0]
    Xgate(pi/2) | q[1]
    
    Rgate(pi/2) | q[0]
    Rgate(pi/2) | q[1]
    
    # Example Oracle 
    # Constant zero in this case
    Xgate(1) | q[1]
    
    Rgate(-pi/2) | q[0]
    
    # Check if the output is zero
    MeasureX | q[0]
\end{lstlisting}
This creates the circuit shown in Figure \ref{fig:DJCircuit} which produces the Wigner function shown in Figure \ref{fig:DJWigner}. 
Note that the example oracle used is constant displacement in the \(x\) direction which results into the Wigner function being concentrated around \(x=0\), indicating that the oracle function is constant one.

\section{Summary} \label{sec:summary}

In this pedagogical paper we have introduced the mathematical tools used in the theory of continuous variable quantum computing (CVQC), from basics of quantum mechanics up to continuous variable quantum algorithms. The reader should now be knowledgeable enough to explore the field starting from the classic reviews \cite{Braunstein2005, Wang2007, Weedbrook2012, Andersen2010} and the algorithms \cite{Pati2003DeutschJozsa, Wagner2012, Pati2000Grover, Lomonaco2002_1, Lomonaco2002_2}. 

There are several important topics which we have not covered in the current version of this paper, however below we make sure to provide the critical references required for getting started with these topics. 

One of them is continuous variable error correction, which can be studied along the same lines as in the case of discrete quantum computing \cite{Braunstein1998PRL, Braunstein1998Nature, Noh2020, Wu2021}. 

The most discussed experimental realizations of CVQC are based on optical devices, where the two quadrature amplitudes of each mode are conjugate continuous variables. Operations are accomplished via various optical elements and homodyne measurement \cite{Killoran2019, Arrazola2021}. Nevertheless, other experimental realizations have also been recently discussed \cite{StrandbergPhD, Hillmann2020}. 

A slight detour from the main course of CVQC, but still closely related field of research is boson sampling (BS) \cite{Aaronson2010}. BS is an active modern field of research, especially on the experimental front \cite{Tillmann2013, Arrazola2021}, and it has been recently used to demonstrate quantum computational advantage \cite{Zhong2020}. 

Gaussian boson sampling (GBS) is a variation of BS which is easier to implement experimentally \cite{Hamilton2017GBS, Kruse2019GBS, Quesada2020}. Besides being yet another candidate of a computational problem which can be used to demonstrate quantum computational advantage, a bunch of applications have been found for GBS as well \cite{Bromley2020}.

\section*{Acknowledgments}
This work has been done as a part of \href{https://qosf.org/qc_mentorship/}{Quantum Computing Mentorship Program} held by Quantum Open Source Foundation (QOSF). S.B. and R.C. are the mentees, H.S. is the mentor.

\printbibliography[heading=bibintoc, title={References}]

\begin{appendices}
\addtocontents{toc}{\protect\setcounter{tocdepth}{-1}}
\section{Uncertainty relations} \label{ap:A}
Generally speaking, the relation \eqref{eq:uncertainty_relation} expresses "uncertainty" in that it necessarily places limits on what we can know about a particle from simultaneous measurements of position and momentum. To conceptually illustrate the Heisenberg's uncertainty principle, consider \(N\) (a very large integer) non-interacting identical particles. This is called an ensemble of particles. Alternatively, you can also consider a single particle in one universe and imagine a lot of identical copies of the universe. Let's measure the coordinate of each particle. We will have \(N\) numbers as a result. The average of these numbers would be the ensemble average of the measured coordinate, and this is what is denoted by \(\bar{x}\). All the other quantities with bars on top of them should be understood in the same manner. If one is looking into a state where \(\Delta x\) is very small, i.e. the coordinate measurement results are narrowly distributed around the average value, we are dealing with a highly localized particle. In this case, Heisenberg's uncertainty principle tells us that the distribution of momentum cannot be narrow, and the particles in the ensemble contain widely distributed values of momentum.

Heisenberg's uncertainty principle is not unique to conjugate continuous variables. In general, for any non commuting continuous variables \(\hat{A}\) and \(\hat{B}\) the following uncertainty relation is true
\begin{equation}
    \Delta A \Delta B \ge \frac{\left\vert \overline{\hat{A}\hat{B}-\hat{B}\hat{A}} \right\vert}{2}, \notag
\end{equation}
where \(\vert\cdot\vert\) means absolute value. In the case if \(\hat{A}\) and \(\hat{B}\) are conjugate variables, their commutator is given by the equation \eqref{eq:commutation} and the Heisnberg's original principle \eqref{eq:uncertainty_relation} is obtained.

\section{Dirac's \(\delta\) function} \label{ap:B}

Dirac's delta function \(\delta(x)\) can be defined by
$$
\int d x f(x) \delta\left(x^{\prime}-x\right)=f\left(x^{\prime}\right).
$$
There is no actual function that does this, although one can think of \(\delta\left(x^{\prime}-x\right)\) as a sort of limit of ordinary functions that vanish when \(x^{\prime}-x\) is not very close to 0 but are very big when \(x^{\prime}-x\) is very close to 0, with the area under the graph of \(\delta\left(x^{\prime}-x\right)\) equal to \(1 .\) The precise way to think of it is that \(\delta\left(x^{\prime}-x\right)\) is a "distribution", where a distribution \(F\) maps nice well behaved functions \(f\) to (complex) numbers \(F[f]\).

A commonly used identity for Dirac's \(\delta\) function is 
\begin{equation}
    \frac{1}{2\pi}\int_{-\infty}^{+\infty}e^{ip(x-x^{\prime})} dp = \delta(x-x^{\prime}) \notag
\end{equation}

\end{appendices}

\end{document}